# Time-frequency Quantum Key Distribution over a Free-Space Optical Link

Jasper Rödiger[1], Nicolas Perlot[1], Oliver Benson[2], Ronald Freund[1]

[1]*Fraunhofer Institute for Telecommunications, Heinrich-Hertz-Institute, Einsteinufer 37, 10587 Berlin, Germany*
[2]*Humboldt-Universität zu Berlin, AG Nanooptik, Newtonstraße 15, 12489 Berlin, Germany*

**Abstract:** We present an implementation of the time-frequency (TF) quantum key distribution (QKD) protocol realized mainly with standard telecommunication components at 1550 nm. TF-QKD is implemented with modulations in time and frequency, namely pulse position modulation (PPM) and frequency shift keying (FSK). The time-frequency uncertainty relation ensures the security of the protocol. We further demonstrate free-space optical QKD transmission over a 388-m distance. Since the QKD setup is single-mode fiber (SMF) -based, precise coupling in and out of the SMFs is crucial. Thus, we implemented an optical tracking system. The QKD signal is wavelength-multiplexed with a strong beacon for optical tracking. Strong spectral filtering is necessary to separate beacon and QKD signal. Together with spatial filtering caused by the SMF most background light is filtered, enabling daylight QKD transmission.

1.   **Introduction**

Secure communication becomes increasingly important and is today mostly addressed by cryptographic algorithms that apply computational problems (e.g. prime factorization or finding discrete logarithms) that would take a classical computer a long time to solve. Important examples of algorithms are asymmetric algorithms, such as RSA [1], Diffie-Hellman [2] or Elliptic Curve Cryptography [3,4] and symmetric algorithms such as AES [5]. However, quantum computers are being developed [6-10], threatening some of those cryptographic schemes, namely by running certain quantum algorithms. The most eminent of those quantum algorithms is Shor's Algorithm for prime factorization and finding discrete logarithms [11-14]. Shor's Algorithm does not threaten symmetric cryptographic algorithms like AES, but these are usually combined with threated asymmetric schemes, e.g. to distribute AES keys.
One method of secure communication, even in the presence of a quantum computer, is quantum key distribution (QKD), which achieved tremendous progress in the last decades [15-18]. To make QKD widely available and cheap, one way is to utilize single-mode fiber (SMF) based off-the-shelf telecommunication components that are in wide use. However, using SMFs also as a transmission channel is sometimes not an option, e.g. because of the cost and time investment necessary for deploying the fibers, or when QKD between mobile partners is desired. On these terms, free-space QKD links are an alternative, albeit with specific technical challenges.
The first and most studied QKD protocol is the BB84 protocol [19]. The photons encode quantum bits (qubits), which are encoded in one of two non-orthogonal bases, with the property that it is not possible to distinguish states in both bases with certainty.
In another approach, time of arrival and frequency of single photons encode the qubits [20,21], namely the time-frequency (TF)-QKD protocol (not to be confused with Twin-Field QKD [22], which also uses the abbreviation TF-QKD in recent years). Here, the photon's arrival time respective to a reference time, or the optical frequency respective to a reference frequency, encodes the qubit. The time-frequency uncertainty relation ensures that the measurement in one basis increases the uncertainty in the other basis and thus deletes information possibly encoded therein. The modulations in the time and frequency domain are robust for fiber and free-space transmissions.
Regarding free-space QKD, a lot of progress has been made in recent years. Here, one important topic is high loss scenarios e.g. in [23] a daylight transmission over 53 km (40 dB loss) in daylight was demonstrated. First satellite QKD experiments have also been carried out, first and foremost with the satellite Micius [28-30].

In the course of this paper, we present an implementation of the TF-QKD protocol, mainly implemented with SMF-based off-the-shelf telecom components. We further report on its performance over a 388 m free-space link in the Berlin city center. Our results demonstrate the feasibility to implement the TF-QKD protocol for free-space QKD with mostly classical-communication components.

## 2. Basics

In TF-QKD two bases are realized by modulations in the time and frequency domain, namely pulse position modulation (PPM) and frequency-shift keying (FSK), in the following referred to as the PPM and FSK basis, respectively. In Fig. 1, the PPM and FSK bases are shown in the time and frequency domain. We consider Fourier-limited Gaussian pulse shapes for all pulses. The PPM (FSK) basis consists of one of two narrow pulses in the time (frequency) domain and of a wide pulse (obeying the Fourier transform) in the frequency (time) domain. A measurement in the PPM (FSK) basis, in other words decreasing the uncertainty there, increases the uncertainty in the FSK (PPM) basis. Thus, assuming appropriate pulse width and separation, the measurement in one basis deletes information encoded in the other basis, analog to the BB84 protocol.

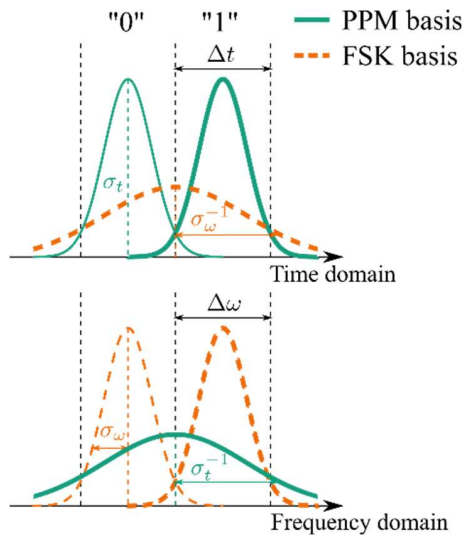

Fig. 1: PPM and FSK basis used for the TF-QKD protocol. The PPM (FSK) basis consists of a narrow Gaussian pulse at one of two positions in the time (frequency) domain and a central broad Gaussian pulse in the frequency domain. Assuming Fourier limited pulses, the width of a pulse in the time domain determines its width in the frequency domain. The measurement of a PPM- (FSK-) encoded pulse in the frequency domain modifies the width of the pulse in the time domain, decreasing the information encoded there

The procedure to distill a secret key is equivalent to the procedure in BB84: Alice randomly prepares symbols that are encoded on photons in one of two states, either in the PPM or FSK basis. Bob measures randomly either in the PPM or FSK basis. Alice and Bob publicly compare the bases they have prepared/measured the photons in and only keep symbols where their basis coincides. From this so-called sifting process, a sifted key is obtained. To estimate the QBER a random subset of the remaining symbols is taken (and discarded afterwards). From this QBER, the amount of information potentially leaked to an eavesdropper is calculated. Using error correction and privacy amplification it is then possible to distill, out of the sifted key, a secret key that is only known to Alice and Bob.

## 3. Implementation

Since our objective is to use existing off-the-shelf telecommunication components, the whole setup is SMF based. For the optical transmission, the telecom wavelength at 1550 nm was chosen due to the low absorption in SMFs and the atmosphere and since many telecommunication components are optimized or only operational at this wavelength.

The implemented setup is shown in Fig. 2. Due to the components at hand, the system's repetition rate is set to 30 MHz, which was close to the maximum switching frequency of the lasers optical frequency. As a test sequence, the symbols and bases were alternated, with the four possible symbols repeating indefinitely. The pulse width and distance parameters, as defined in [31], can be found in Table 1.

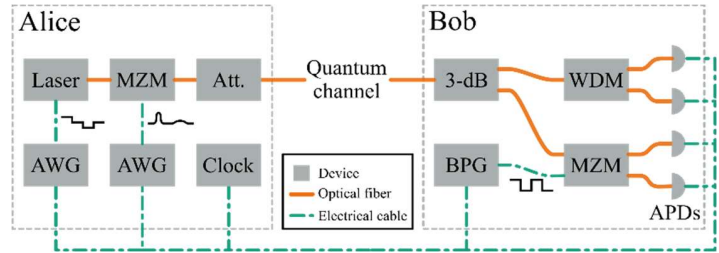

Fig. 2: Scheme of the implemented TF-QKD setup. Alice: A tunable distributed brag reflector (DBR) laser, whose optical frequency is controlled by an arbitrary waveform generator (AWG), sends out continuous wave (CW) light. A Mach-Zehnder modulator (MZM) shapes the Gaussian pulses controlled by a second AWG. The AWGs are synchronized by a clock. The signal is attenuated and sent to Bob. Bob: A 3-dB coupler served as a passive basis choice. A 12.5-GHz wavelength division (de-)multiplexer (WDM) sorts the incoming photons according to their optical frequency and forwards them to one of two avalanche photon diodes (APDs). A double output (DO) MZM, controlled by a bit pattern generator and synchronized with the clock, sorts the incoming photons according to their time of arrival and forwards them to one of two APDs. The APDs are gated with the gating signal also being synchronous with the clock.

**Table 1. Pulse parameters**

| Time-domain | | Frequency-domain | |
| --- | --- | --- | --- |
| quantity | value | quantity | value |
| $\sigma_t$ | 97 ps | $\sigma_\omega$ | 3.6 GHz[a] |
| $\sigma_\omega^{-1}$ | 281 ps | $\sigma_t^{-1}$ | 10.3 GHz[a] |
| $\Delta t$ | 977 ps | $\Delta\omega$ | 35.7 GHz |

[a]Calculated regarding the respective pulse within the time domain

In Alice's part of the setup, a tunable distributed-Bragg-reflector (DBR) laser emits a continuous-wave (CW) signal with the desired optical frequency and a laser-linewidth below 5 MHz. To modulate the signal in the frequency domain the phase of the DBR laser is controlled by an arbitrary waveform generator (AWG) that generates a voltage pattern with three levels, namely V0 for PPM pulses and V1 or V2 for FSK pulses (see Fig. 3 a)).

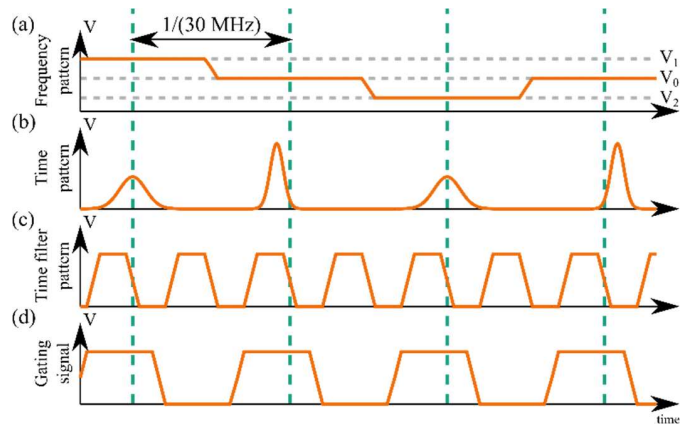

Fig. 3: Scheme of the patterns produced by arbitrary waveform generators (AWGs) and bit pattern generators (BPGs). (a) pattern controlling the optical frequency; (b) pattern used for modulating the pulses in the time domain; (c) pattern used for the double output Mach-Zehnder modulator (DOMZM) for filtering in the pulse in the time domain. (d) gating pattern for the avalanche photon-diodes (APDs).

The CW laser light is forwarded to a 40-Gbit/s Mach-Zehnder modulator (MZM), which creates the desired pulses in the time domain and thus, assuming Fourier-limited pulses, determines the pulse width in the frequency domain, which is large compared to the laser linewidth. The MZM is driven by a 34 GSa/s AWG that shapes the desired Gaussian pulses with 6-bit resolution (see Fig. 3 b)). Both AWGs are synchronized by a central clock, also synchronizing Alice and Bob. The laser pulses are attenuated by an optical attenuator to an average photon number of 0.5 photons per pulse. This photon number is suggested for a decoy-state protocol [32] in [33]. Implementing decoy states would be straightforward by either using the existing or an additional MZM. However, in the presented setup decoy states are not implemented. The photons are forwarded to Bob over a free-space link as described below.

In Bob's setup, a 3-dB-coupler randomly forwards the received photons to either the PPM or the FSK basis and thus functions as a passive basis choice. Since the time resolution of the InGaAs avalanche photon-diodes (APDs) at hand is too low to discriminate consecutive pulses with a high enough fidelity, the PPM-basis measurement was realized with a double-output (DO)MZM used as a time-filter. The DOMZM is controlled by a voltage pattern generated by a bit-pattern generator (BPG)). The BPG is also synchronized with the clock. Technical length restrictions of the BPG pattern (see Fig. 3 c)) required the pattern to be repeated with a faster rate than the 30-MHz symbol rate. The two DOMZM outputs are connected each to an APD, which then click, depending on the time of arrival.

The FSK-basis measurement consists of a 12.5 GHz wavelength division (de)multiplexer (WDM) that sorts the photons according their optical frequency. The used optical frequencies (following $V_{0-2}$) are chosen according to the WDM. The two outputs of the WDM are connected to two additional APDs. The length of a QKD-transmission measurement was bound by the storage capacity of the measurement electronics and thus depends on the number of photons measured by the APDs. For the below presented free-space transmission, a measurement could be up to about 100 minutes long.

In comparison to APDs sensitive at lower wavelengths, InGaAs APDs have a relatively high noise level, thus in the presented setup, they are gated, as is common practice [34]. The gate width, namely, for how long an APD is sensitive every cycle, is an important optimizing parameter of QKD systems. It influences the dark counts as well as the amount of collected signal photons (as will be discussed in Section 4). A suitable gating signal is generated from the clock signal (see Fig. 3 d)). The phase of the patterns shown in Fig. 3 can be changed to synchronize Alice and Bob and to account for the transmission delay.

The detection signals of the APDs are measured and evaluated after each transmission process by comparing the measurement with the sent test pattern and basis choice. From that comparison, the sifted key rate and quantum bit error rate (QBER) are determined.

The free-space optical link is established between two buildings in the Berlin city center, see Fig. 4. To double the distance and to have Alice and Bob conveniently next to each other, a remotely controllable mirror is positioned at one of the buildings. Because the QKD system is entirely SMF based, it is necessary to couple the optical signal in and out of the SMF efficiently. Here, two optical antennas developed in-house [35] are deployed for Alice and Bob. Single-mode beams are reciprocal, even in a turbulent medium [36]. Reciprocity is exploited with bidirectional beams for tracking: each antenna measures the received beam with a quadrant detector and corrects tip and tilt accordingly by means of a fine steering mirror, thus maintaining a pointing error sustainably below 10 µrad.

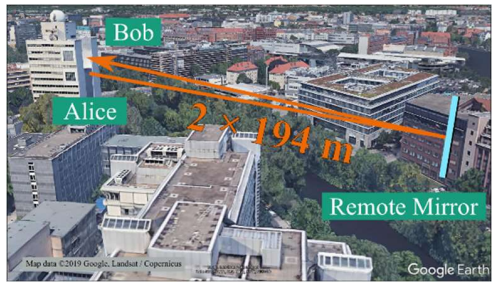

Fig. 4: Free-space testbed in the Berlin city center between two buildings. Alice and Bob are localized in Fraunhofer Heinrich Hertz Institute (HHI) main building (left-hand side). In a HHI site branch (right-hand-side) a remotely controllable mirror is installed to double the link distance and make the test operations more convenient.

Since the quantum signal is very weak, bidirectional beacon beams with an optical frequency 400 GHz away from the quantum signal are superimposed to the unidirectional quantum signal. Thus, the beacon and quantum signal need to be separated sufficiently to prevent additional noise, accomplished by the implementation of multiple add-drop WDM filters and circulators, as described in [35]. With three filters, 150 dB of isolation is achieved while only inducing 2 dB of additional loss. The spectrally immediate proximity of quantum signal and beacon also shows that spectrally efficient multiplexing with classical communication is possible. The losses acquired over the free-space link varied between 10 and 16 dB at clear sky weather conditions.

## 4. Experimental Results

With the described optical tracking system, it is possible to maintain a stable optical link. Further, the WDM filtering, together with the spatial filtering of the SMF field, removes enough background light to allow QKD in daylight.

Since a general security proof is still missing for the TF-QKD protocol, we apply the analysis conducted in [31] regarding an intercept/resend attack to estimate the secret key rate. The sifted key rate, QBER and secret key can be seen in Fig. 5 as a function of the gate width of the APDs. As expected, it can be observed that a larger gate width increases the QBER by increasing the amount of dark counts, but also increases the sifted key rate by increasing the number of collected photons. A QBER as low as 4 % was achieved. However, at a different gate width of 0.48 ns, the maximum secret key rate was 8.9 kbit/s with a QBER of around 7 %.

## 5. Conclusion

We presented an implementation of the TF-QKD protocol with mainly SMF-based telecom components at the 1550 nm telecom wavelength. With suitable tracking antennas, capable of efficiently coupling light in and out of SMFs, and multiplexing the QKD signal with a beacon spectrally just 400 GHz apart, transmission over a free-space optical link of 388 m is demonstrated. While it is not shown in the presented work, the beacon could very well be used e.g. as a classical communication channel. The fact that the beacon wavelength is so close to the QKD wavelength shows, that a high portion of spectral range is available for multiplexing of other classical channels, especially, since Raman scattering is not significant with the few meters of optical fiber integrated in the setup.

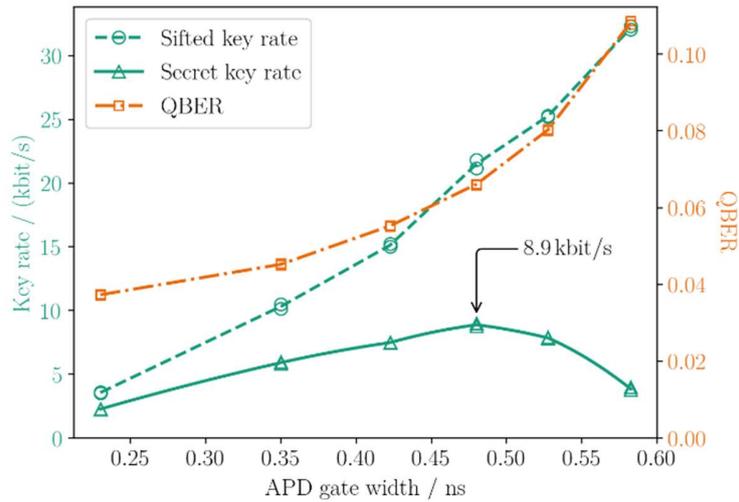

Figure 5: QKD transmission over the 388 m link. The measured sifted key rate and quantum bit error rate (QBER) are shown, together with the calculated secret key rate (see [31] for details on the calculation) plotted over the gate width of the APDs as an optimization parameter. A secret key rate of up to 8.9 kbit/s was achieved.

For the TF-QKD protocol, a general security proof is still missing. However, full security proofs for other protocols, also based on the time-frequency (or time-energy) uncertainty relation do exist [37-39], hinting that a general proof is in principle possible for the TF-QKD protocol. The presented secret key rate can thus be seen as an estimate. A secret key rate of up to 8.9 kbit/s is possible under the assumed intercept/resend attack. We show that sufficient filtering makes daylight transmission possible.

Due to the components at hand, the repetition rate of the presented system was set to 30 MHz. In principle, it would be possible to increase this repetition rate. One bottleneck in the presented setup is the tuning frequency achievable with the DBR laser. Multiple modulated lasers could dissolve this bottleneck. Further, the gating frequency of the implemented APDs cannot exceed 100 MHz. APDs with gating frequencies in the GHz range [40-42] or superconducting nanowire single-photon detectors [43,-44] without the need for gated operation, could increase the key rate.

Summarizing, we showed that the implemented protocol has a lot of potential for free-space optical QKD in daylight especially, since multiplexing with spectrally close (just 400 MHz apart) classical communication channels is very much feasible.

**Funding.** This work has been funded by the German Research Foundation (DFG) within the CRC 787, project C2.

**Disclosures.** The authors declare no conflicts of interest.

# References

1. R. L. Rivest, A. Shamir, and L. Adleman, "A method for obtaining digital signatures and public-key cryptosystems," Communications of the ACM, vol. 21, no. 2, pp. 120–126, 1978.
2. W. Diffie and M. Hellman, "New directions in cryptography," IEEE transactions on Information Theory, vol. 22, no. 6, pp. 644–654, 1976.
3. V. S. Miller, "Use of elliptic curves in cryptography," in Conference on the theory and application of cryptographic techniques. Springer, 1985, pp. 417–426.
4. N. Koblitz, "Elliptic curve cryptosystems," Mathematics of computation, vol. 48, no. 177, pp. 203–209, 1987.
5. J. Nechvatal, E. Barker, L. Bassham, W. Burr, M. Dworkin, J. Foti, and E. Roback, "Report on the development of the advanced encryption standard (AES)," Journal of Research of the National Institute of Standards and Technology, vol. 106, no. 3, p. 511, 2001.
6. M. Steffen, D. P. DiVincenzo, J. M. Chow, T. N. Theis, and M. B. Ketchen, "Quantum computing: An IBM perspective," IBM Journal of Research and Development, vol. 55, no. 5, pp. 13–1, 2011.
7. N. M. Linke, D. Maslov, M. Roetteler, S. Debnath, C. Figgatt, K. A. Landsman, K. Wright, and C. Monroe, "Experimental comparison of two quantum computing architectures," *Proceedings of the National Academy of Sciences*, vol. 114, no. 13, pp. 3305–3310, 2017.
8. K. Svore, A. Geller, M. Troyer, J. Azariah, C. Granade, B. Heim, V. Kliuchnikov, M. Mykhailova, A. Paz, and M. Roetteler, "Q#: Enabling scalable quantum computing and development with a high-level DSL," *Proceedings of the Real World Domain Specific Languages Workshop* 2018. ACM, 2018, p. 7.
9. S. Boixo, S. V. Isakov, V. N. Smelyanskiy, R. Babbush, N. Ding, Z. Jiang, M. J. Bremner, J. M. Martinis, and H. Neven, "Characterizing quantum supremacy in near-term devices," *Nature Physics*, vol. 14, no. 6, p. 595, 2018.
10. C. Neill, P. Roushan, K. Kechedzhi, S. Boixo, S. V. Isakov, V. Smelyanskiy, A. Megrant, B. Chiaro, A. Dunsworth, K. Arya, R. Barends, B. Burkett, Y. Chen, Z. Chen, A. Fowler, B. Foxen, M. Giustina, R. Graff, E. Jeffrey, T. Huang, J. Kelly, P. Klimov, E. Lucero, J. Mutus, M. Neeley, C. Quintana, D. Sank, A. Vainsencher, J. Wenner, T. C. White, H. Neven, and J. M. Martinis, "A blueprint for demonstrating quantum supremacy with superconducting qubits," *Science*, vol. 360, no. 6385, pp. 195–199, 2018.
11. P. W. Shor, "Algorithms for quantum computation: Discrete logarithms and factoring," *Proceedings 35th annual symposium on foundations of computer scienc*e. IEEE, 1994, pp. 124–134.
12. E. Martín-López, A. Laing, T. Lawson, R. Alvarez, X.-Q. Zhou, and J. L. O'Brien, "Experimental realization of Shor's quantum factoring algorithm using qubit recycling," *Nature Photonics*, vol. 6, no. 11, p. 773, 2012.
13. E. Lucero, R. Barends, Y. Chen, J. Kelly, M. Mariantoni, A. Megrant, P. O'Malley, D. Sank, A. Vainsencher, J. Wenner, T. White, Y. Yin, A. N. Cleland, and John M. Martinis, "Computing prime factors with a Josephson phase qubit quantum processor," *Nature Physics*, vol. 8, no. 10, p. 719, 2012.
14. T. Monz, D. Nigg, E. A. Martinez, M. F. Brandl, P. Schindler, R. Rines, S. X. Wang, I. L. Chuang, and R. Blatt, "Realization of a scalable shor algorithm," *Science*, vol. 351, no. 6277, pp. 1068–1070, 2016.
15. A. Boaron, G. Boso, D. Rusca, C. Vulliez, C. Autebert, M. Caloz, M. Perrenoud, G. Gras, F. Bussières, M-J Li, D. Nolan, A. Martin, and H. Zbinden, "Secure Quantum Key Distribution over 421 km of Optical Fiber," *Phys. Rev. Lett.* 121, 190502, 2018
16. N. T. Islam, C. C. W. Lim, C. Cahall, J. Kim, and D. J. Gauthier, "Provably secure and high-rate quantum key distribution with time-bin qudits." *Science advances*, 3(11), e1701491, 2017
17. C. Agnesi, M. Avesani, L. Calderaro, A. Stanco, G. Foletto, M. Zahidy A. Scriminich, F. Vedovato, G Vallone, and P. Villoresi, "Simple Quantum Key Distribution with qubit-based synchronization and a self-compensating polarization encoder," *Optica*, 7(4), 284-290, 2020
18. S. Pirandola, U. L. Andersen, L. Banchi, M. Berta, D. Bunandar, R. Colbeck, D. Englund, T. Gehring, C. Lupo, C. Ottaviani, J. Pereira, M. Razavi, J. S. Shaari, M. Tomamichel, V. C. Usenko, G. Vallone, P. Villoresi, P. Wallden, "Advances in Quantum Cryptography*," arXiv preprint* quantum-ph/1906.01645, 2019
19. H. Ch. Bennett and G. Brassard, "Quantum cryptography: public key distribution and coin tossing," *Conf. on Computers, Systems and Signal Processing (Bangalore, India, Dec. 1984)*, pp. 175–9, 1984.
20. Z. Chang-Hua, P. Chang-Xing, Q. Dong-Xiao, G. Jing-Liang, C. Nan, and Y. Yun-Hui, "A new quantum key distribution scheme based on frequency and time coding," *Chinese Physics Letters*, vol. 27, no. 9, p. 090301, 2010.
21. S. Yelin and B. C. Wang, "Time-frequency bases for bb84 protocol," *arXiv preprint*, quant-ph/0309105, 2003.
22. M. Lucamarini, Z. L. Yuan, J. F. Dynes, and A. J. Shields, "Overcoming the rate–distance limit of quantum key distribution without quantum repeaters," Nature, 557(7705), 400-403, 2018.
23. S. Liao, H. L. Yong, C. Liu, G. L. Shentu, D. D. Li, J. Lin, H. Dai, S.-Q. Zhao, B. Li, J.-Y. Guan, W. Chen, Y.-H. Gong, Y. Li, Z.-H. Lin, G.-S. Pan, J. S. Pelc, M. M. Fejer, W.-Z. Zhang, W.-Y. Liu, J. Yin, J.-G. Ren, X.-B. Wang, Q. Zhang, C.-Z. Peng, and J.-W. Pan, "Long-distance free-space quantum key distribution in daylight towards inter-satellite communication," *Nature Photonics*, 11(8), 509-513, 2017
24. Y.-H. Gong, K.-X. Yang, H.-L. Yong, J.-Y. Guan, G.-L. Shentu, C. Liu, F.-Z. Li, Y. Cao, J. Yin, S.-K. Liao, J.-G. Ren, Q. Zhang, C.-Z. Peng, and J.-W. Pan, "Free-space quantum key distribution in urban daylight with the SPGD algorithm control of a deformable mirror," *Optics express*, 26(15), 18897-18905, 2018
25. J. Jin, J.-P. Bourgoin, R. Tannous, S. Agne, C. J. Pugh, K. B. Kuntz, B. L. Higgins, and T. Jennewein, "Genuine time-bin-encoded quantum key distribution over a turbulent depolarizing free-space channel," *Optics Express*, 27(26), 37214-37223, 2019

26. H. Chen, J. Wang, B. Tang, Z. Li, B. Liu, and S. Sun, "Field demonstration of time-bin reference-frame-independent quantum key distribution via an intracity free-space link," *Optics Letters*, 45(11), 3022-3025, 2020
27. M. Avesani, L. Calderaro, M. Schiavon, A. Stanco, C. Agnesi, A. Santamato, M. Zahidy, A. Scriminich, G. Foletto, G. Contestabile, M. Chiesa, D. Rotta, M. Artiglia, A. Montanaro, M. Romagnoli, V. Sorianello, F. Vedovato, G. Vallone, P. Villoresi "Full daylight quantum-key-distribution at 1550 nm enabled by integrated silicon photonics," *arXiv preprint,* arXiv:1907.10039, 2019
28. S.-K. Liao, W.-Q. Cai, W.-Y. Liu, L. Zhang, Y. Li, J.-G. Ren, J. Yin, Q. Shen, Y. Cao, Z.-P. Li, F.-Z. Li, X.-W. Chen, L.-H. Sun, J.-J. Jia, J.-C. Wu, X.-J. Jiang, J.-F. Wang, Y.-M. Huang, Q. Wang, Y.-L. Zhou, L. Deng, T. Xi, L. Ma, T. Hu, Q. Zhang, Y.-A. Chen, N.-L. Liu, X.-B. Wang, Z.-C. Zhu, C.-Y. Lu, R. Shu, C.-Z. Peng, J.-Y. Wang, and J.-W. Pan "Satellite-to-ground quantum key distribution," *Nature*, 549(7670), 43-47, 2017
29. J. Yin, Y. Cao, Y.-H. Li, J.-G. Ren, S.-K. Liao, L. Zhang, W.-Q. Cai, W.-Y. Liu, B. Li, H. Dai, M. Li, Y.-M. Huang, L. Deng, L. Li, Q. Zhang, N.-L. Liu, Y.-A. Chen, C.-Y. Lu, R. Shu, C.-Z. Peng, J.-Y. Wang, and J.-W. Pan, "Satellite-to-ground entanglement-based quantum key distribution," *Physical review letters*, 119(20), 200501, 2017
30. S.-K. Liao, W.-Q. Cai, J. Handsteiner, B. Liu, J. Yin, L. Zhang, D. Rauch, M. Fink, J.-G. Ren, W.-Y. Liu, Y. Li, Q. Shen, Y. Cao, F.-Z. Li, J.-F. Wang, Y.-M. Huang, L. Deng, T. Xi, L. Ma, T. Hu, L. Li, N.-L. Liu, F. Koidl, P. Wang, Y.-A. Chen, X.-B. Wang, M. Steindorfer, G. Kirchner, C.-Y. Lu, R. Shu, R. Ursin, T. Scheidl, C.-Z. Peng, J.-Y. Wang, A. Zeilinger, J.-W. Pan, „Satellite-relayed intercontinental quantum network," *Physical review letters*, 120(3), 030501, 2018
31. J. Rödiger, N. Perlot, R. Mottola, R. Elschner, C.-M. Weinert, O. Benson, and R. Freund, "Numerical assessment and optimization of discrete-variable time-frequency quantum key distribution*," Phys. Rev. A,* vol. 95, p. 052312, May 2017.
32. W.Y. Hwang, "Quantum key distribution with high loss: toward global secure communication," *Physical Review Letters*, 91(5), 057901, 2003
33. Y. Zhang, I. B. Djordjevic, and M. A. Neifeld, "Weak-coherent-state based time-frequency quantum key distribution," *Journal of Modern Optics*, vol. 62, no. 20, pp. 1713–1721, 2015.
34. Z. L.Yuan, B. E. Kardynal, A. W. Sharpe, and A. J. Shields, . "High speed single photon detection in the near infrared," *Applied Physics Letters*, 91(4), 041114, 2007.
35. N. Perlot, J. Rödiger, and R. Freund, "Single-mode optical antenna for high-speed and quantum communications," *Photonic Networks; 19th ITG-Symposium*. VDE, 2018.
36. J. H. Shapiro, and A. L. Puryear. "Reciprocity-enhanced optical communication through atmospheric turbulence—Part I: Reciprocity proofs and far-field power transfer optimization," *Journal of Optical Communications and Networking* 4.12 : 947-954, 2012.
37. Z. Zhang, J. Mower, D. Englund, F. N. Wong, and J. H. Shapiro, "Unconditional security of time-energy entanglement quantum key distribution using dual-basis interferometry," *Physical Review Letters*, vol. 112, no. 12, p. 120506, 2014.
38. N. T. Islam, C. C. W. Lim, C. Cahall, J. Kim, and D. J. Gauthier, "Provably secure and high-rate quantum key distribution with time-bin qudits," *Science Advances*, vol. 3, no. 11, p. e1701491, 2017.
39. T. Zhong, H. Zhou, R. D. Horansky, C. Lee, V. B. Verma, A. E. Lita, A. Restelli, J. C. Bienfang, R. P. Mirin, T. Gerrits, S. W. Nam, F. Marsili, M. D. Shaw, Z. Zhang, L. Wang, D. Englund, G. W. Wornell, J. H. Shapiro, and F. N. C. Wong , "Photonefficient quantum key distribution using time–energy entanglement with high-dimensional encoding," *New Journal of Physics*, vol. 17, no. 2, p. 022002, 2015.
40. Y.Liang, Y. Chen,, Z. Huang, G. Bai, M. Yu, and H. Zeng, (2016), "Room-temperature single-photon detection with 1.5-GHz gated InGaAs/InP avalanche photodiode," *IEEE Photonics Technology Letters*, 29(1), 142-145.
41. M. Ren, X. Gu, Y. Liang, W. Kong, E. Wu, G. Wu, and H. Zeng, (2011), "Laser ranging at 1550 nm with 1-GHz sine-wave gated InGaAs/InP APD single-photon detector," *Optics Express*, 19(14), 13497-13502.
42. Liang, Y., Fei, Q., Liu, Z., Huang, K., and Zeng, H. (2019), "Low-noise InGaAs/InP single-photon detector with widely tunable repetition rates," *Photonics Research*, 7(3), A1-A6.
43. E. A. Dauler, M. E. Grein, A. J. Kerman, F. Marsili, S. Miki, S. W. Nam, M. D. Shaw, H. Terai, V. B. Verma, and T. Yamashita, "Review of superconducting nanowire single-photon detector system design options and demonstrated performance," *Optical Engineering*, vol. 53, no. 8, pp. 081 907– 081 907, 2014.
44. C. M. Natarajan, M. G. Tanner, and R. H. Hadfield, "Superconducting nanowire single-photon detectors: physics and applications," *Superconductor Science and Technology*, vol. 25, no. 6, p. 063001, 2012.